\documentclass[8.5pt,twoside,twocolumn]{article}
\usepackage[super,sort&compress,comma]{natbib}
\usepackage{mhchem}
\usepackage{times,mathptmx}
\usepackage{sectsty}
\usepackage{balance}

\usepackage{graphicx} 
\usepackage{lastpage}
\usepackage[format=plain,justification=raggedright,singlelinecheck=false,font=small,labelfont=bf,labelsep=space]{caption}
\usepackage{fancyhdr}
\usepackage{color}
\newcommand{\si}{ESI\dag}

\newcommand{\ib}[1]{{\color{black}#1}}

\begin{document}

\thispagestyle{plain}
\fancypagestyle{plain}{
\renewcommand{\headrulewidth}{1pt}}
\renewcommand{\thefootnote}{\fnsymbol{footnote}}
\renewcommand\footnoterule{\vspace*{1pt}%
\hrule width 3.4in height 0.4pt \vspace*{5pt}}
\setcounter{secnumdepth}{5}

\makeatletter
\def\subsubsection{\@startsection{subsubsection}{3}{10pt}{-1.25ex plus -1ex minus -.1ex}{0ex plus 0ex}{\normalsize\bf}}
\def\paragraph{\@startsection{paragraph}{4}{10pt}{-1.25ex plus -1ex minus -.1ex}{0ex plus 0ex}{\normalsize\textit}}
\renewcommand\@biblabel[1]{#1}
\renewcommand\@makefntext[1]%
{\noindent\makebox[0pt][r]{\@thefnmark\,}#1}
\makeatother
\renewcommand{\figurename}{\small{Fig.}~}
\sectionfont{\large}
\subsectionfont{\normalsize}

\fancyfoot{}
\fancyfoot[RO]{\footnotesize{\sffamily{1--\pageref{LastPage} ~\textbar  \hspace{2pt}\thepage}}}
\fancyfoot[LE]{\footnotesize{\sffamily{\thepage~\textbar\hspace{3.45cm} 1--\pageref{LastPage}}}}
\fancyhead{}
\renewcommand{\headrulewidth}{1pt}
\renewcommand{\footrulewidth}{1pt}
\setlength{\arrayrulewidth}{1pt}
\setlength{\columnsep}{6.5mm}
\setlength\bibsep{1pt}

\newcommand{\alt}{\raisebox{-0.3ex}{$\stackrel{<}{\sim}$}}
\newcommand{\agt}{\raisebox{-0.3ex}{$\stackrel{>}{\sim}$}}

\twocolumn[
  \begin{@twocolumnfalse}
\noindent\LARGE{\textbf{
Electrochemical setup --- a unique chance to simultaneously control
orbital energies and vibrational properties of single-molecule junctions with unprecedented efficiency\dag
}}
\vspace{0.6cm}

\noindent\large{\textbf{Ioan B\^aldea $^{\ast}$
\textit{$^{a\ddag}$}
}}\vspace{0.5cm}


\noindent \textbf{\small{Published: Phys.~Chem.~Chem.~Phys.~2014, {\bf 16}, 25942-25949, DOI: 10.1039/C4CP04316B}}
\vspace{0.6cm}

\noindent 
\normalsize{Abstract:\\
Impressive advances in nanoscience permit nowadays to manipulate single molecules and broadly control many of their properties. Still, tuning the molecular charge and vibrational properties of single molecules embedded in nanojunctions in broad ranges escaped so far to an efficient control. By combining theoretical results with recent experimental data, we show that, under electrochemical control, it is possible to continuously drive a redox molecule (viologen) between almost perfect oxidized and reduced states. This yields an unprecedentedly efficient control on both vibrational frequencies and the surface-enhanced Raman scattering (SERS) intensities. The broad tuning achieved under electrochemical control by varying the overpotential (``gate potential'') within experimentally accessible ranges contrasts to the case of two-terminal setups that require high biases, which real nanojunctions cannot withstand. The present study aim at stimulating concurrent transport and SERS measurements in electrochemical setup. This may open a new avenue of research that is not accessible via two-terminal approaches for better understanding the transport at nanoscale.

$ $ \\  

}
\vspace{0.5cm}
 \end{@twocolumnfalse}
  ]


\footnotetext{\textit{$^{a}$~Theoretische Chemie, Universit\"at Heidelberg, Im Neuenheimer Feld 229, D-69120 Heidelberg, Germany.}}
\footnotetext{\dag~Electronic supplementary information (ESI) available. See DOI: 10.1039/C4CP04316B}
\footnotetext{\ddag~E-mail: ioan.baldea@pci.uni-heidelberg.de.
Also at National Institute for Lasers, Plasmas, and Radiation Physics, Institute of Space Sciences,
Bucharest, Romania}
\section{Introduction}
\label{sec:intro}
Despite important advances in the last decades, molecular electronics remains
confronted with a series of difficulties. Many of them result from
unsatisfactory characterization of a molecule under \emph{in situ} conditions.
Properties of a molecule embedded in a nanojunction in a current-carrying state
may differ from its equilibrium properties in a manner reminding differences
between biological cells in \emph{in vivo} and \emph{in vitro} situations.
Fundamental processes at nanoscale and tailoring molecular devices
for practical applications can be
better understood if transport data can be correlated with other molecular properties
obtained from independent measurements of a different kind.
Vibrational properties deduced via surface-enhanced Raman spectroscopy (SERS)
belong to this category, since they can provide valuable complementary information
needed for a better \emph{in situ} characterization.

Recent concurrent SERS \cite{LeRu:12}
and ohmic conductance studies on single-molecule junctions
represent significant attempts in this direction.
Since the first demonstration that SERS can be employed to study molecular junctions,
\cite{Tao:06e} there is a continuing interest to utilize this technique
in molecular electronics.\cite{Tao:07a,Natelson:08a,Natelson:08c,Ioffe:08,Liu:11,Wandlowski:11b,Konishi:13a}
The electrodes of molecular junctions can act as highly efficient
plasmonic antennas.\cite{Muehlschlegel:05}
The enormous local electric field
(especially close to the sharp tip of a scanning tunneling microscope \cite{Liu:11}
or a nanoparticle \cite{Tian:10})
can have a dramatic effect on the surface-enhanced Raman scattering
from molecules in junctions.
Correlating SERS data with simultaneously acquired
conductance data provides important
evidence on the chemical identity of the active molecule,
on how it bonds to electrodes, and on experimental conditions
(\emph{e.g.}, solvent, sample treatment, \emph{etc}).
At low biases corresponding to a linear
transport
regime, these refer
to properties of the molecule (linked to electrodes but) at equilibrium
(fluctuation-dissipation theorem).
In particular, they refer to a molecule in a given charge (often neutral) state.

Applying higher biases
$V_b$
on a molecular junction
may change molecular properties,
also including the charge of the molecule, and this change can reflect itself in a change of its vibrational properties.
A recent SERS study on fullerene-based
electromigrated junctions \cite{Natelson:14}
found that vibrational frequencies are significantly shifted under applied bias. Companion
density functional theory (DFT) calculations indicated that the observed frequency shifts are
inconsistent with a simple vibrational Stark effect, but they can result from a bias-driven
change of the electronic charge of the molecule.\cite{Natelson:14}
By applying
realistic
source-drain
voltages
$V_b$
in a two-terminal setup, it is only a
\emph{very partial} reduction
(in contradistinction to a complete reduction of the molecular species, corresponding to changing the molecular charge by an entire electron)
that can be achieved
(\emph{cf.}\
Ref.~\citenum{Natelson:14} and
{\figurename}\ref{fig:ivna}).

The theoretical results reported in this paper, obtained by combining
a model study backed by companion
quantum chemical calculations and existing experimental data, \cite{Wandlowski:08}
demonstrate that an SERS study in electrochemical scanning
tunneling microscopy (EC-STM) setup
can provide valuable information complementary
to that obtained via a
single-molecule
transport study.
The EC-STM approach exploits the flexibility of a three-terminal setup: both the
bias $V_b \equiv V_t - V_s$ between the STM-tip ($t$) and substrate ($s$)
and the overpotential $\eta \equiv V_{eq} - V_s$ \cite{Tao:96,Wandlowski:08}
can be independently controlled. For viologen-based junctions,\cite{Wandlowski:08}
the equilibrium potential $V_{eq} \simeq -0.46$\,V.
Charge transport through several redox systems in EC-STM setup
has been experimental studied
\cite{Alessandrini:05,Alessandrini:06,VisolyFisher:06,Wandlowski:08,ZhangKuznetsov:08,Artes:12b}
but corresponding SERS experiments are missing.
The advantages of this setup introduced in
Tao's seminal work,\cite{Tao:96}
which enables a practically complete reduction/oxidation of the molecular species,
for concurrent SERS and transport studies will be emphasized.
\section{Methods}
\label{sec:methods}
To demonstrate the usefulness of a concurrent SERS and transport study through
a redox unit embedded in a single-molecule junction in a three-terminal electrochemical setup,
we will consider viologen-based molecular junctions, which were investigated in a comprehensive
experimental study.\cite{Wandlowski:08}

The core of this molecule consists of a redox-active 4,4$^\prime$-bipyridinum dication (44BPY$^{++}$).
The parent (neutral bipyridine 44BPY$^{0}$) molecule embodied in nanojunctions
formed the object of numerous transport,\cite{Tao:03,Venkataraman:09b,Baldea:2013b} transport-related
\cite{Baldea:2012i,Baldea:2013a,Baldea:2013c,Baldea:2014a} and
SERS \cite{Joo:04,Liu:11,Konishi:13a} studies.
This redox-active molecule was used, \emph{e.g.}, as backbone in self-assembled monolayers \cite{Long:92,Sagara:98,Alvarado:05}
and in various functional materials.\cite{Lahav:00,Akiyama:05}
It is a showcase redox molecule; the first oxidation-reduction process
44BPY$^{++} \rightleftharpoons$ 44BPY$^{+\bullet}$, which will be examined below,
is completely reversible in bulk solutions.\cite{Wandlowski:08}

High-level quantum chemical calculations at the density functional theory (DFT) level
using the Becke's three-parameter hybrid functional
B3LYP and basis sets of triple-zeta quality augmented with diffuse functions (aug-cc-pVTZ)
as implemented in the GAUSSIAN 09 package \cite{g09} have been performed
for geometry optimizations and for obtaining the Raman spectra of the oxidized and reduced viologen core
(dication 44BPY$^{++}$ and radical cation 44BPY$^{+\bullet}$, respectively), which are the
charge species that contribute to the measured current \cite{Wandlowski:08} in the specific case considered in this paper.

In addition to the validation against the experimental transport data, we have also microscopically validate
the Newns-Anderson model employed below 
by calculating the lowest electron affinities at the EOM-CCSD
(equation-of-motion coupled cluster singles and doubles) 
level,\cite{Baldea:2014c} which represents the
quantum chemistry state-of-the-art for molecules of this size; see the \si.

To compute the vibrational frequencies and Raman scattering intensities of
a molecular junction under bias ($V_b \neq 0, \eta \neq 0$),
we have resorted to an interpolation weighting method described in the {\si},
wherein the weight is expressed in terms of the $V_b$- and $\eta$-dependent
LUMO occupancy.
\section{Results}
\label{sec:results}
\begin{figure}[h!]
$ $\\[5ex]
\centerline{\includegraphics[width=0.4\textwidth,angle=0]{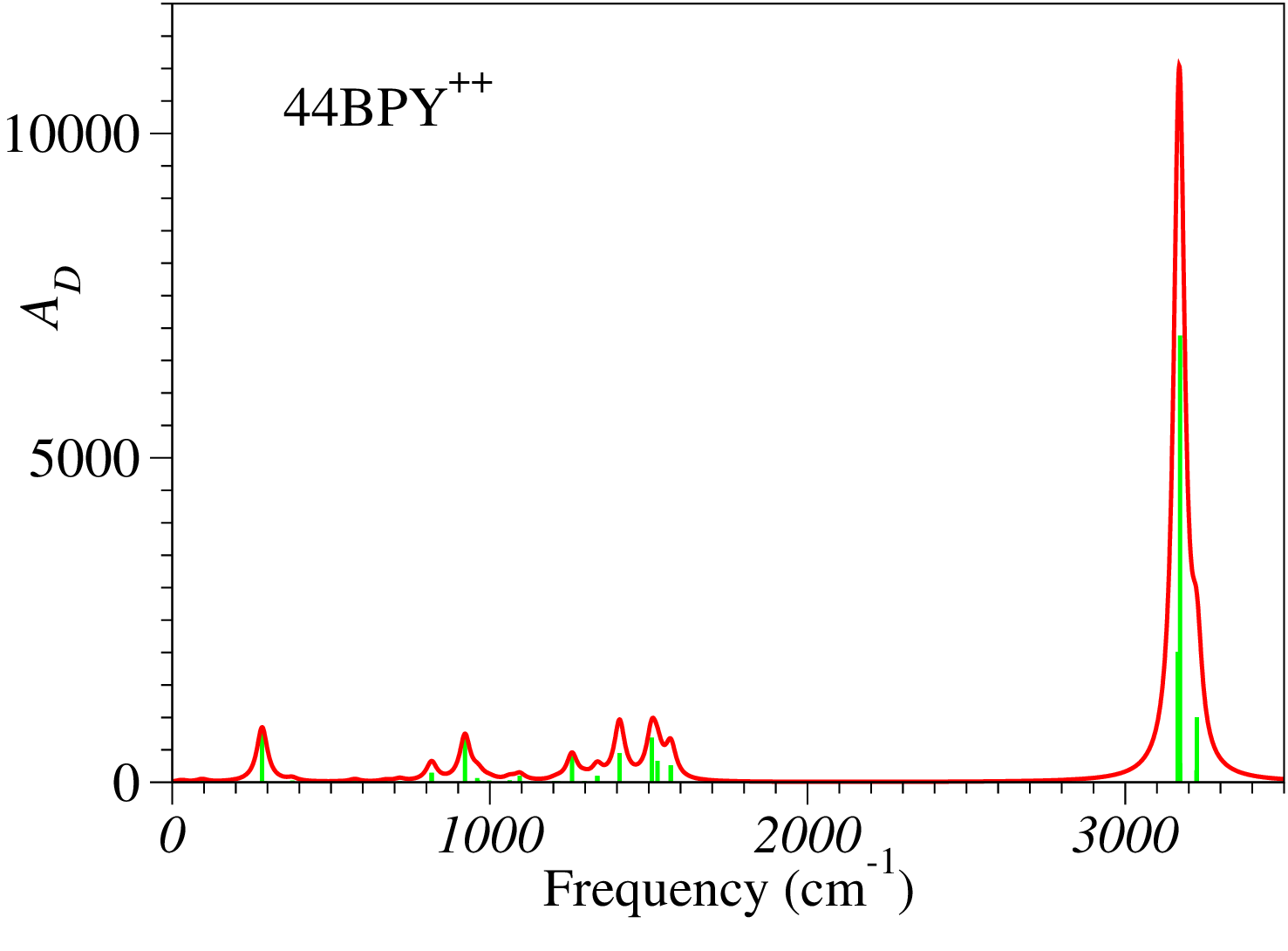}}
$ $\\[0ex]
\caption{Raman scattering activities (spectral lines in {\AA}$^4$/a.m.u.) of the dication 44BPY$^{++}$ in acetonitrile.
The envelope (red line) has been obtained by convoluting the computed spectral lines (green spikes) with Lorentzian functions of half-width 20\,cm$^{-1}$.}
\label{fig:raman-dication}
\end{figure}

\begin{figure}[h!]
$ $\\[5ex]
\centerline{\includegraphics[width=0.4\textwidth,angle=0]{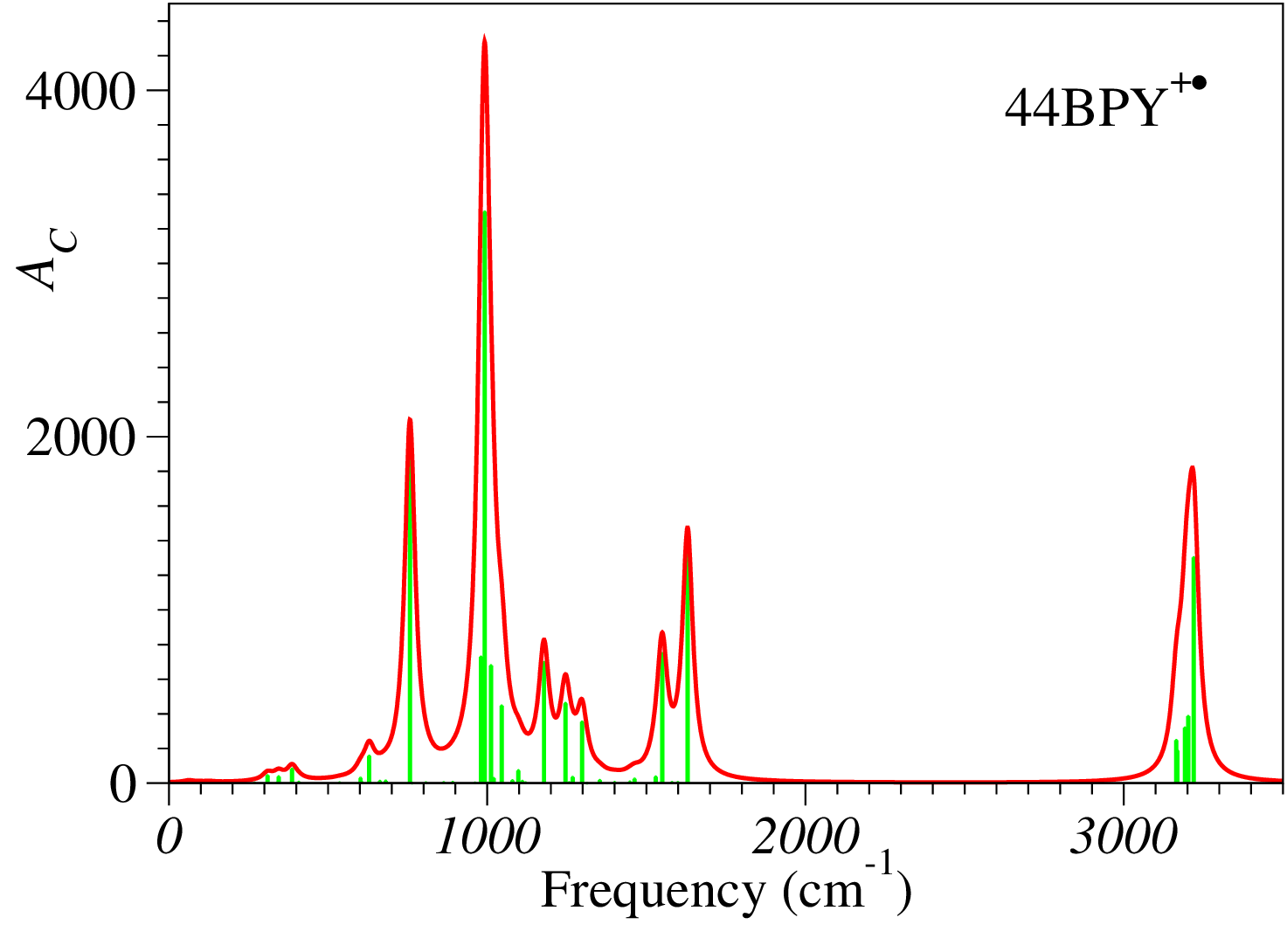}}
$ $\\[0ex]
\caption{Raman scattering activities (spectral lines in {\AA}$^4$/a.m.u.) of the cation 44BPY$^{+\bullet}$ 
in acetonitrile.
The envelope (red line) has been obtained by convoluting the computed spectral lines 
(green spikes) with Lorentzian functions of half-width 20\,cm$^{-1}$.}
\label{fig:raman-cation}
\end{figure}

The Raman spectra of the dication (44BPY$^{++}$)
and cation (44BPY$^{+\bullet}$) species in acetonitrile
computed as described in {\si} are presented in
Table S1 
and
{\figurename}\ref{fig:raman-dication},
{\figurename}\ref{fig:raman-cation},
{\figurename}S1, 
{\figurename}S2, 
{\figurename}S3, 
and
{\figurename}S4. 
(Throughout, label S refers to {\si}).
These results reveal notable differences between vibrational properties of the two 
different redox charge species.
Both vibrational frequencies 
and Raman scattering intensities significantly depend on the charging state.

For SERS observability, the important issue is whether a transport setup
permits to broadly control the molecular charge, enabling
a continuously switching between the dicationic (44BPY$^{++}$)
and cationic (44BPY$^{+\bullet}$)
species. They correspond to an oxidized ($n_l = 0$)
and a reduced ($n_l = 1$) LUMO, respectively.
As a central point of the present analysis, we have used the transport data
of Ref.~\citenum{Wandlowski:08} to show that the EC-STM transport setup does enable this switching.

\begin{figure}[h!]
$ $\\[5ex]
\centerline{\includegraphics[width=0.4\textwidth,angle=0]{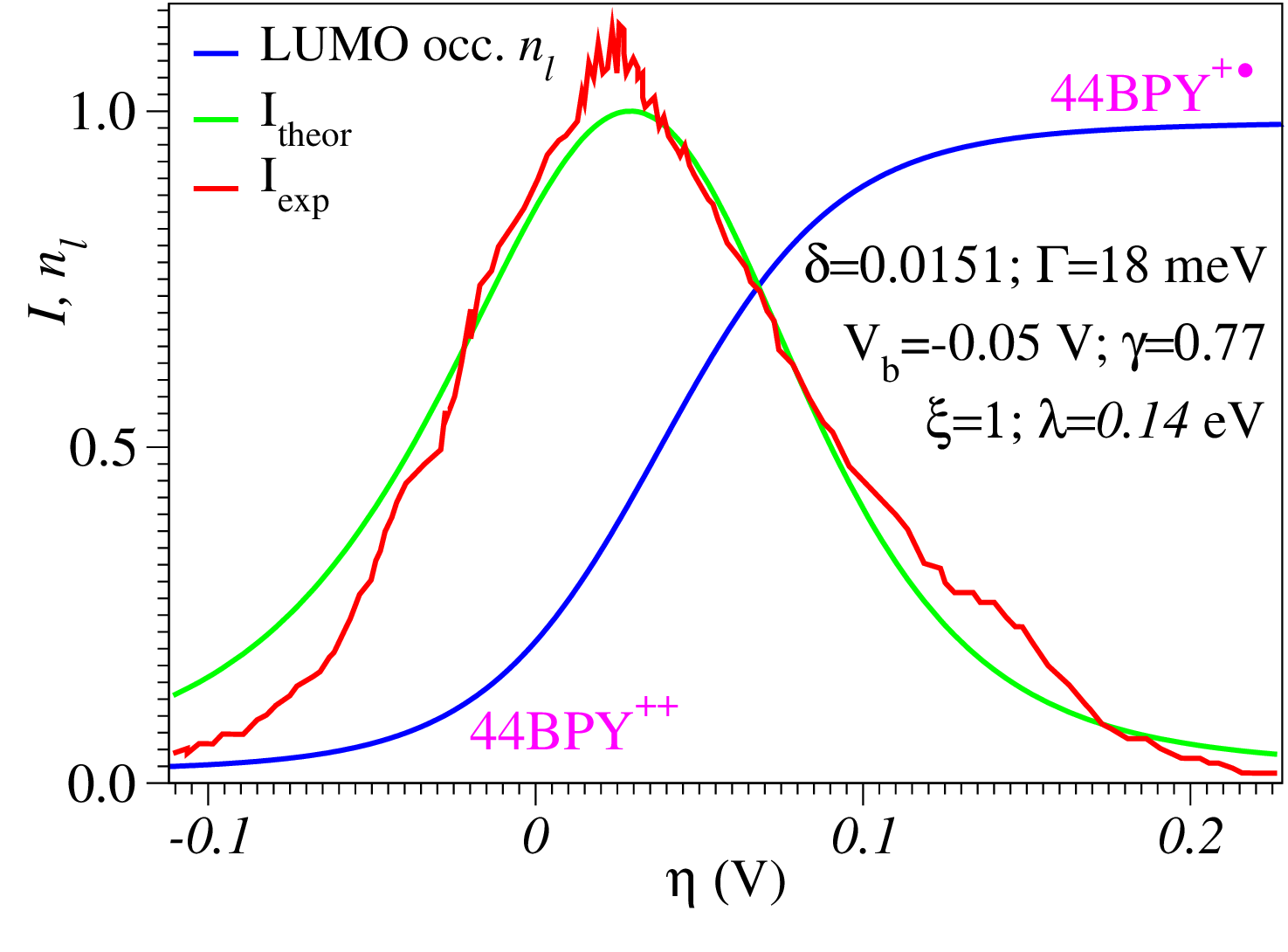}}
$ $\\[0ex]
\caption{The present theoretical model, described in 
detail elsewhere,\cite{Baldea:2013d,1-step-vs-2-steps} is able
to reproduce the
currents $I$ measured by varying the overpotential $\eta$ in constant bias
mode for single-molecule junctions
based on viologen \cite{Wandlowski:08,I_enh} and to show that the LUMO occupancy $n_l$ can
be continuously tuned between $n_l \approx 0$ and  $n_l \approx 1$, which correspond to almost perfect
oxidized and reduced states, respectively. Relevant details on the model and
the parameters $\xi$, $\lambda$, $\gamma$, $\Gamma$, and $\delta$ are given
in {\si}. The experimental current (red curve) presented here was obtained by digitizing 
{\figurename}8A of ref.~\citenum{Wandlowski:08}, 
where the value of the preset current is $I_{T0}=0.1$\,nA; notice that
$E_b=+0.05$\,V of ref.~\citenum{Wandlowski:08} corresponds 
in the present notation to $V_{b} = -0.05$\,V.  
(Currents scaled such that the maximum theoretical 
current is equal to unity.)
}
\label{fig:igna}
\end{figure}

\begin{figure}[h!]
$ $\\[5ex]
\centerline{\includegraphics[width=0.4\textwidth,angle=0]{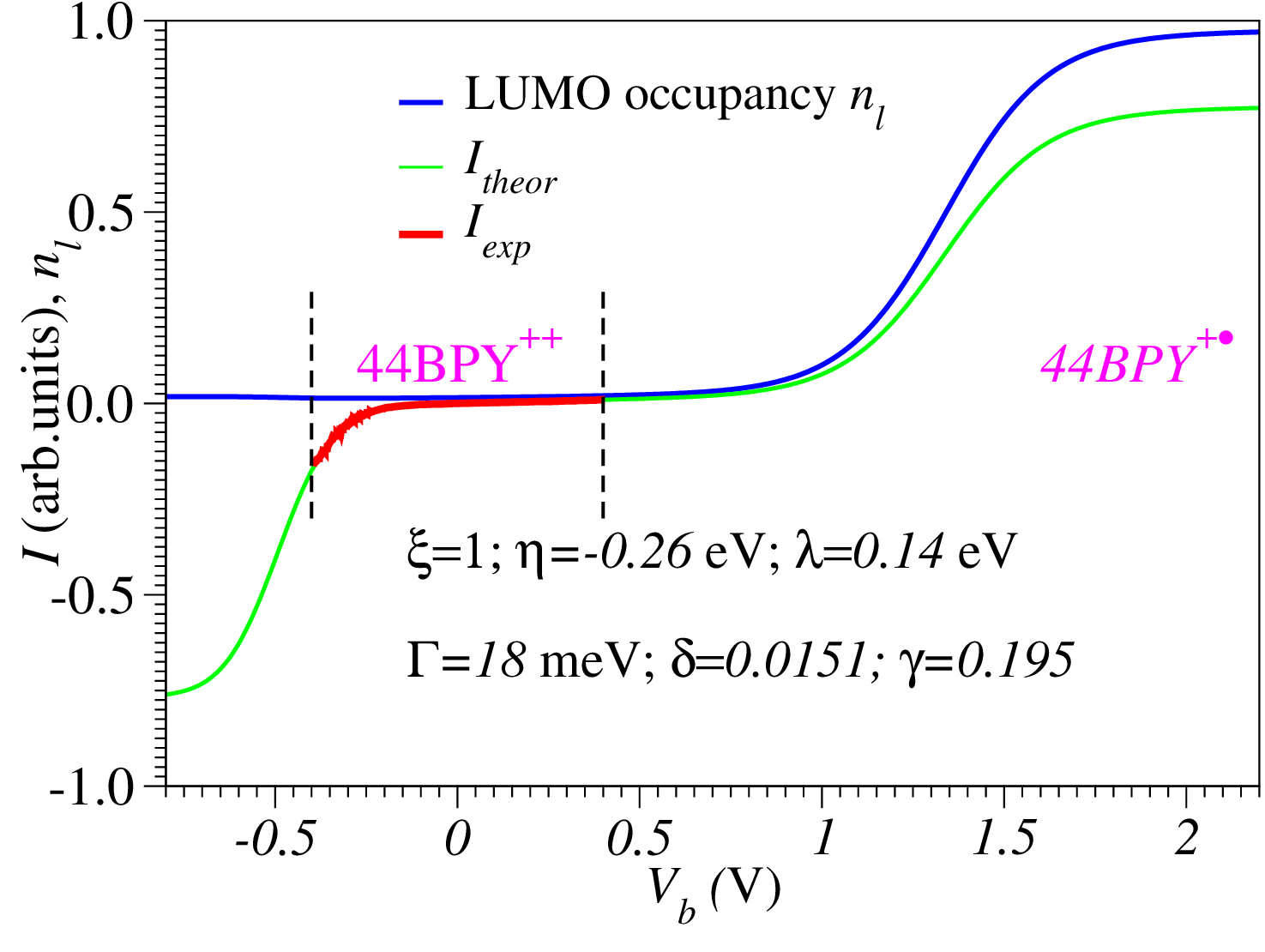}}
$ $\\[0ex]
\caption{The Newns-Anderson model with reorganization, described in detail elsewhere,\cite{Baldea:2013d,1-step-vs-2-steps} is able to reproduce
the experimental curve $I-V$ curve measured for single-molecule junctions based on viologen.
\cite{Wandlowski:08} In the bias ranged sampled in experiment \cite{Wandlowski:08} (depicted by the
red experimental curve and indicated by the two black vertical dashed lines)
the LUMO occupancy is negligible ($n_l \approx 0$). This demonstrates that
an effective reduction cannot be achieved by using source-drain voltages in the experimentally
accessed $V_b$-range. Relevant details on the model and
the parameters $\xi$, $\lambda$, $\gamma$, $\Gamma$, and $\delta$ are given
in {\si}. 
}
\label{fig:ivna}
\end{figure}

In an EC-STM setup, transport data can be acquired in two basic modes: constant bias
and variable bias modes.
In constant bias mode, the
STM-tip ($t$) and substrate ($s$) potentials $V_{t,s}$
are varied such that $V_b = V_t - V_s$ is kept constant.
In variable mode, $V_b$ is varied at constant substrate potential $V_s$ ($\eta = const$).
As visible in {\figurename}\ref{fig:igna} and
{\figurename}\ref{fig:ivna}, the theoretical curves 
successfully reproduce the experimental currents \cite{Wandlowski:08}
measured in both aforementioned modes. 
\ib{Because the Newns-Anderson model utilized to obtain these theoretical curves
has been discussed in detail elsewhere, only a few relevant details are given below 
and in the \si.

An aspect worth to mention is the LUMO position.
The LUMO energy utilized in the transport calculations
lies at $- e V_{eq}\simeq 0.46$\,eV above electrodes' equilibrium Fermi energy, in agreement
with the experimental data.\cite{Wandlowski:08}
Noteworthy, this value, which implicitly enters the definition of the overpotential
(see Sec.~\ref{sec:intro} and, for example, the Supporting Information of ref.~\citenum{Baldea:2013d}),
corresponds to a LUMO energy of the embedded molecule, is different
from that of the isolated molecule. Schemes to disentangle this energy difference
in contributions with clear physical origin have been discussed
recently.\cite{Baldea:2013b,Baldea:2014e}

As another particularly relevant detail, we mention the substantial asymmetry 
of the molecule-electrode couplings
$\delta \equiv \Gamma_{t}/\Gamma$, where $\Gamma \equiv \left(\Gamma_{s} + \Gamma_{t}\right)/2$, 
of the viologen-based EC-STM junctions.\cite{Wandlowski:08} The value $\delta \ll 1$ 
(see legends of {\figurename}\ref{fig:ivna} and {\figurename}\ref{fig:igna}), or alternatively
$\Gamma_t \ll \Gamma_s$,
realistically accounts for the experimental setup of ref.~\citenum{Wandlowski:08};
the viologen molecule is chemically bound to the substrate ($s$) but not to the STM tip ($t$).
Albeit substantial, this asymmetry is not so pronounced as recently found for
azurin-based EC-STM junctions.\cite{Baldea:2013d}
}

Once validating the transport model, we have employed it to compute the bias dependent
LUMO occupancy $n_l$, which is also shown in {\figurename}\ref{fig:igna} and
{\figurename}\ref{fig:ivna}.

\begin{figure}[h!]
$ $\\[5ex]
\centerline{\includegraphics[width=0.4\textwidth,angle=0]{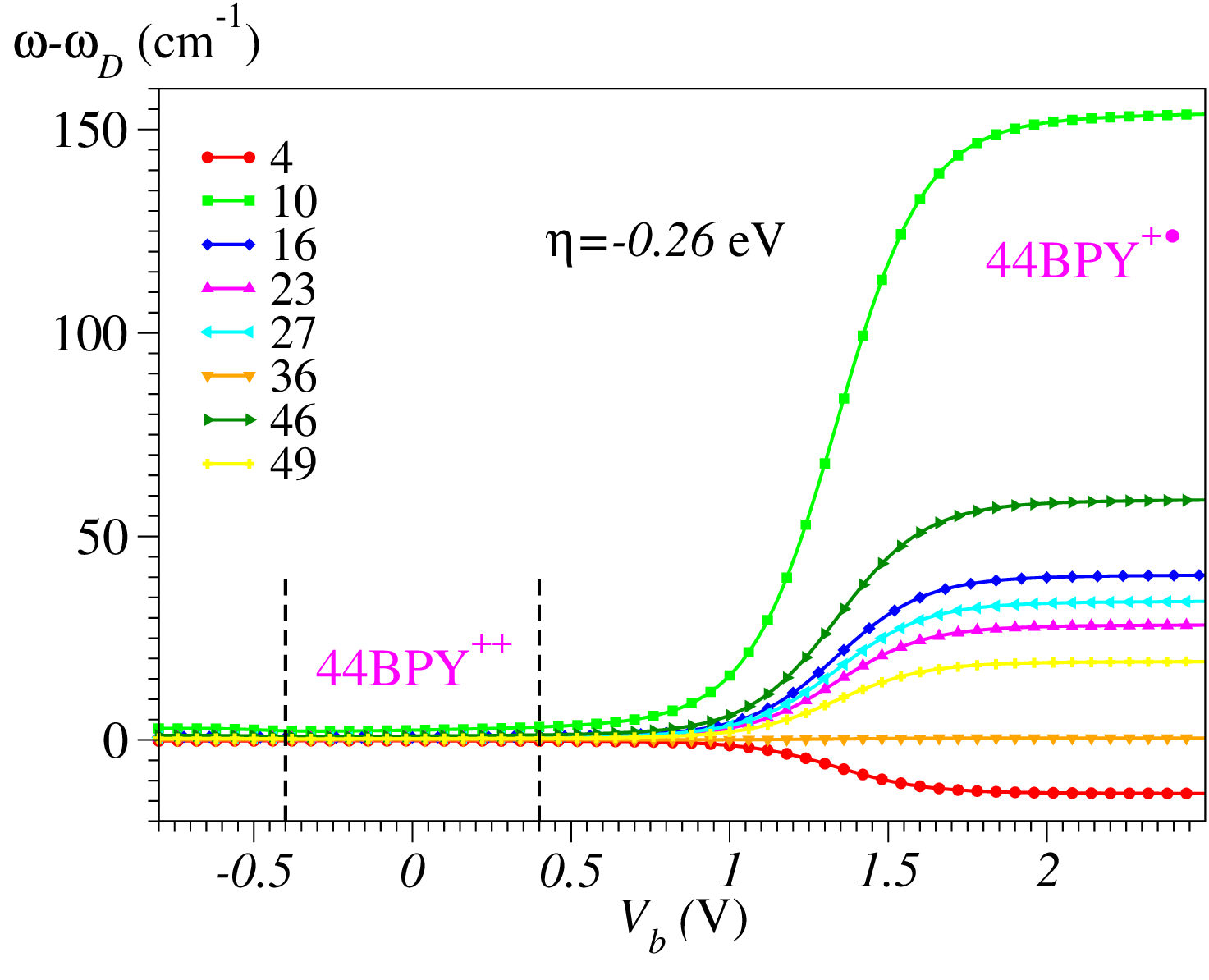}}
$ $\\[2ex]
\centerline{\includegraphics[width=0.4\textwidth,angle=0]{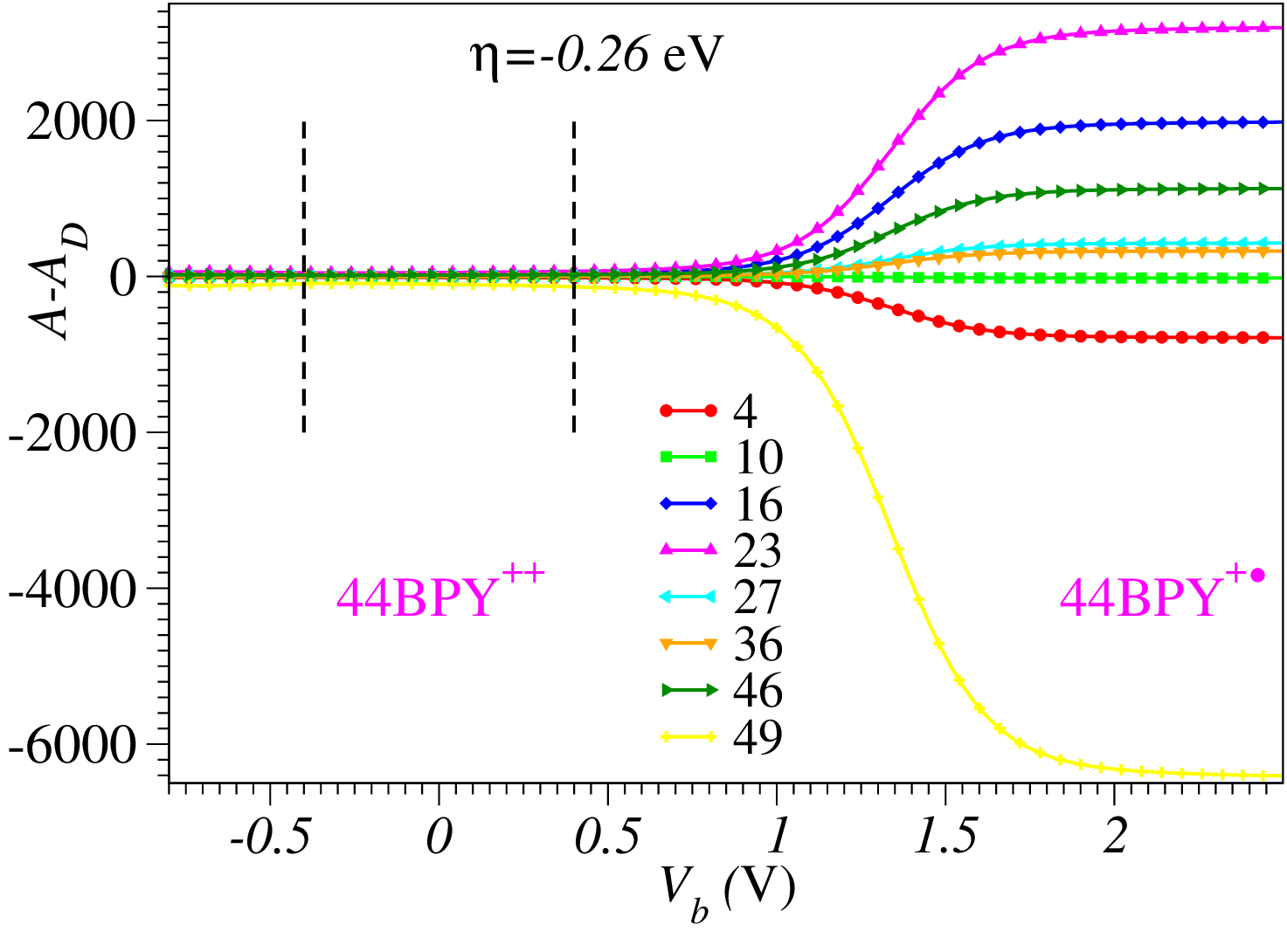}}
\caption{Dependence on the source-drain bias $V_b$ of the changes in frequencies and Raman scattering activities (in {\AA}$^4$/a.m.u.)
of several representative vibrational modes specified in the legend.
As visible, in the range sampled in experiment,\cite{Wandlowski:08} which is indicated by the two black vertical dashed lines,
the vibrational properties do not significantly vary with $V_b$.}
\label{fig:freq-A-vs-V}
\end{figure}

\begin{figure}[h!]
$ $\\[5ex]
\centerline{\includegraphics[width=0.4\textwidth,angle=0]{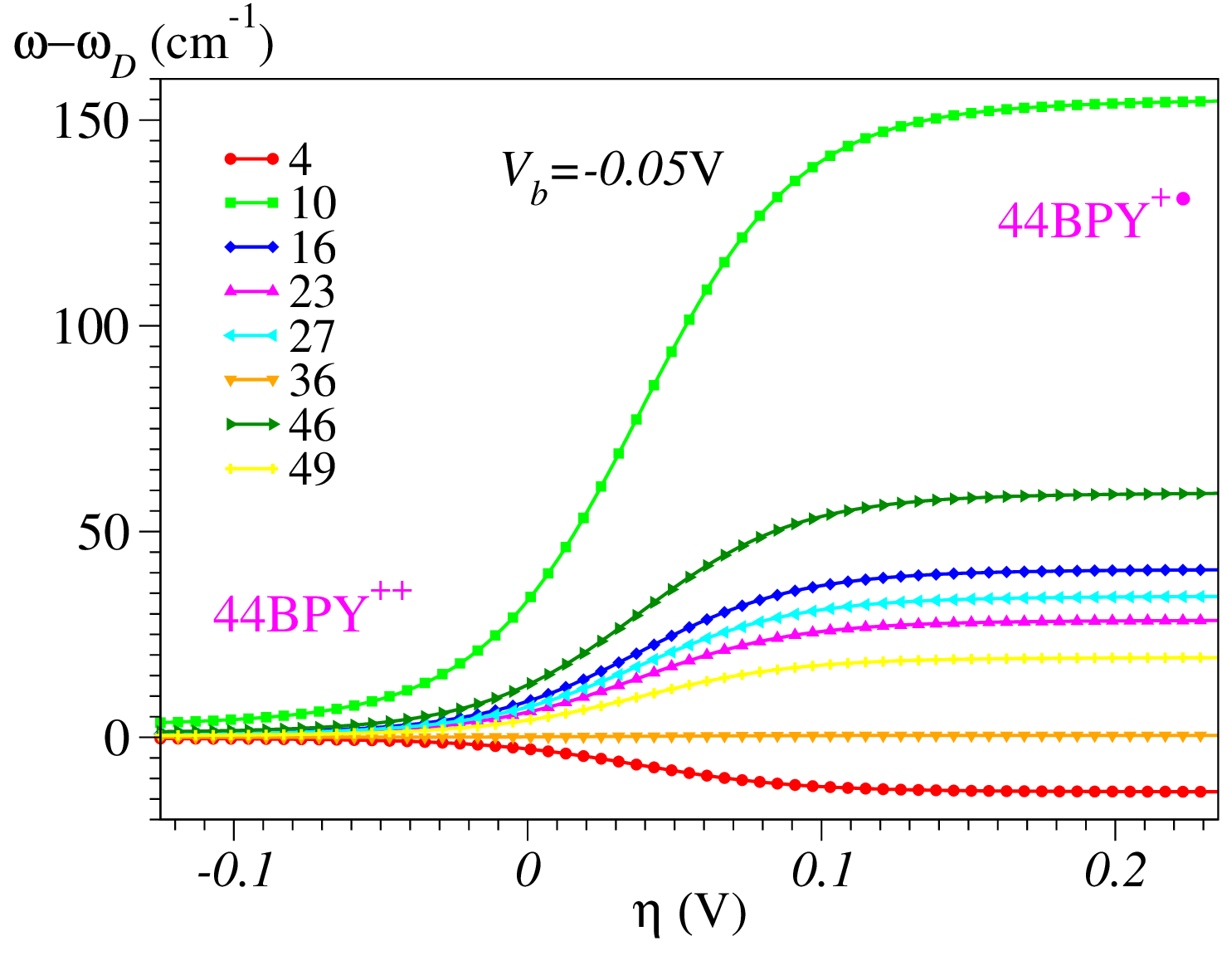}}
$ $\\[2ex]
\centerline{\includegraphics[width=0.4\textwidth,angle=0]{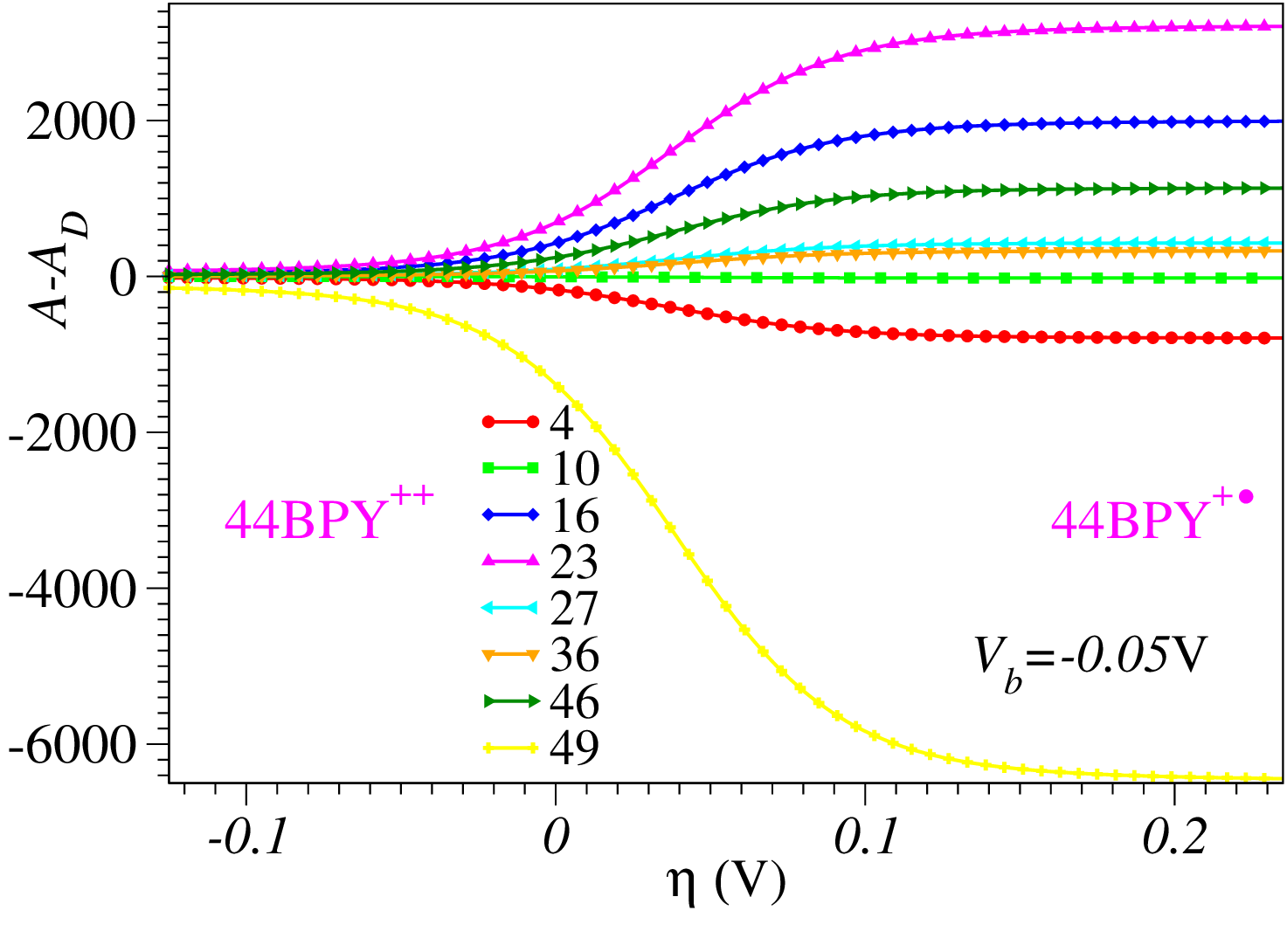}}
\caption{Dependence on the overpotential $\eta$ of the changes in frequencies and Raman scattering activities (in {\AA}$^4$/a.m.u.)
of several representative vibrational modes specified in the legend.}
\label{fig:freq-A-vs-eta}
\end{figure}

As alternative to other approaches to SERS in biased molecular junctions,\cite{Galperin:14a}
to estimate the vibrational frequencies $\omega_{\nu}$ and the
Raman scattering intensities $A_{\nu}$ of the various modes $\nu$
we adopt here an interpolation method described in {\si}.
The $V_b$- and $\eta$-dependencies of $\omega_{\nu}$ and $A_{\nu}$
follow from those of $n_l$. They are depicted in
{\figurename}\ref{fig:freq-A-vs-V} and {\figurename}\ref{fig:freq-A-vs-eta}.
The dependence $n_l = n_l(\eta)$ shown in {\figurename}\ref{fig:igna} represents a key point of the present analysis.
\section{Discussion}
\label{sec-disc}
{\figurename}\ref{fig:igna} and
{\figurename}\ref{fig:ivna}
show that, \emph{in principle},
the charge of a molecule embedded in a biased
EC-STM junction can be controlled both in constant bias mode and variable bias mode.
However, as far as the bias ranges (that can be) accessed in experiment \cite{Wandlowski:08} are concerned,
there is an important quantitative
difference between the two operating modes.

Within the whole $V_b$-range
that has been sampled in experiment ($\vert V_b\vert < 0.4$\,V \cite{Wandlowski:08})
the LUMO occupancy is negligible ($n_l < 0.02$, \emph{cf.}\
{\figurename}\ref{fig:ivna}).
This corresponds to an almost perfect oxidized 
44BPY$^{++}$ state.
As shown in
{\figurename}\ref{fig:freq-A-vs-V},
this yields a negligible $V_b$-dependence of the
vibrational frequencies and Raman scattering intensities.
A significant change in LUMO occupancy, also accompanied by significant variations of the vibrational properties
(\emph{cf.}\
{\figurename}\ref{fig:freq-A-vs-V}),
can only be achieved at substantially larger $V_b$'s,
which molecular junctions could hardly withstand. This is a general feature of the
off-resonant tunneling in two-terminal setup.

{\figurename}\ref{fig:igna} depicts a totally different situation. By varying the overpotential $\eta$
within the range accessed in experiment,\cite{Wandlowski:08}
the LUMO occupancy can be continuously tuned between an almost perfect oxidized state
($n_l \agt 0$, 44BPY$^{++}$) and an almost perfect reduced state ($n_l \alt 1$, 44BPY$^{+\bullet}$).
Because constant bias experiments can practically sample the whole range $0 < n_l < 1$,
the values $\omega_{\nu}$ and $A_{\nu}$ can continuously cover
the values of the various ($\nu$) vibrational modes corresponding to the dicationic and cationic species.
The complete list of $\omega_{\nu}$'s and $A_{\nu}$'s for  44BPY$^{++}$ and 44BPY$^{+\bullet}$
is presented in Table S1 
$\eta$- and $V_b$-dependencies of several representative vibrational modes are shown in
{\figurename}\ref{fig:freq-A-vs-V} and
{\figurename}\ref{fig:freq-A-vs-eta}. In contrast to the insignificant impact of $V_b$ ({\figurename}\ref{fig:freq-A-vs-V}),
$\eta$-driven variations $\omega_{\nu} = \omega_{\nu}(\eta)$ and $A_{\nu} = A_{\nu}(\eta)$
like those depicted there are substantial ({\figurename}\ref{fig:freq-A-vs-eta}).
Frequencies and intensities of the various Raman active modes are affected in different ways.
For some modes, the state of charge only has a weak impact on the frequency while the intensity is strongly affected
and \emph{vice versa}.
The frequency of mode 10 (ring out-of-plane deformation) exhibits the highest frequency change:
$\omega_{10, C} - \omega_{10, D} = 157.6$\,cm$^{-1}$. Although the Raman activity decreases by one order of magnitude
upon reduction (\emph{cf.}~Table S1), 
the small
intensity of this mode would probably be challenging for experimentalists. At the other extreme, the LUMO reduction
only yields a decrease in frequency of mode 49 (CH stretch) amounting to $\omega_{49, C} - \omega_{49, D} = 19.7$\,cm$^{-1}$.
However, the Raman intensity of this mode (which is the highest for the dication,
\emph{cf.}~Table S1) 
is diminished by a factor $A_{49, D}/A_{49, C} \approx 22$. This makes it a good candidate to be monitored in experiments.
The other modes shown in {\figurename}\ref{fig:freq-A-vs-V} and
{\figurename}\ref{fig:freq-A-vs-eta}
are those identified in experimental Raman spectra of the radical anion 44BPY$^{-\bullet}$;\cite{Kassab:96}
see {\si} for further details.
Here we only note that mode 64 is related to the so-called quinoidal distortion,
corresponding to a shortening of the inter-ring C-C bond and of the C-C bond parallel to it,
and a lengthening of the C-C bond between them as well as of the C-N bond.
\cite{Kassab:96,Baldea:2012i}

\begin{figure}[h!]
$ $\\[5ex]
\centerline{\includegraphics[width=0.4\textwidth,angle=0]{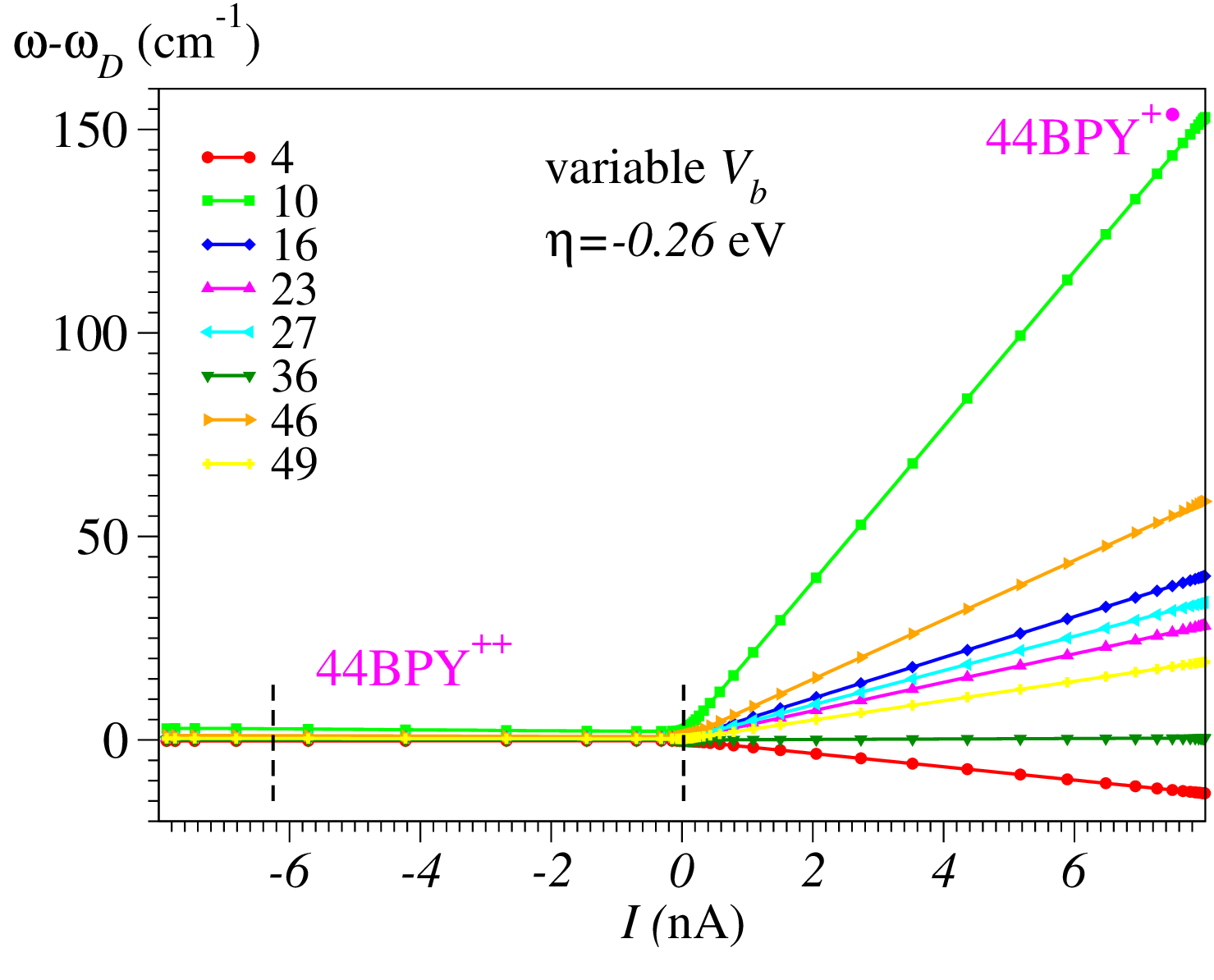}}
$ $\\[2ex]
\centerline{\includegraphics[width=0.4\textwidth,angle=0]{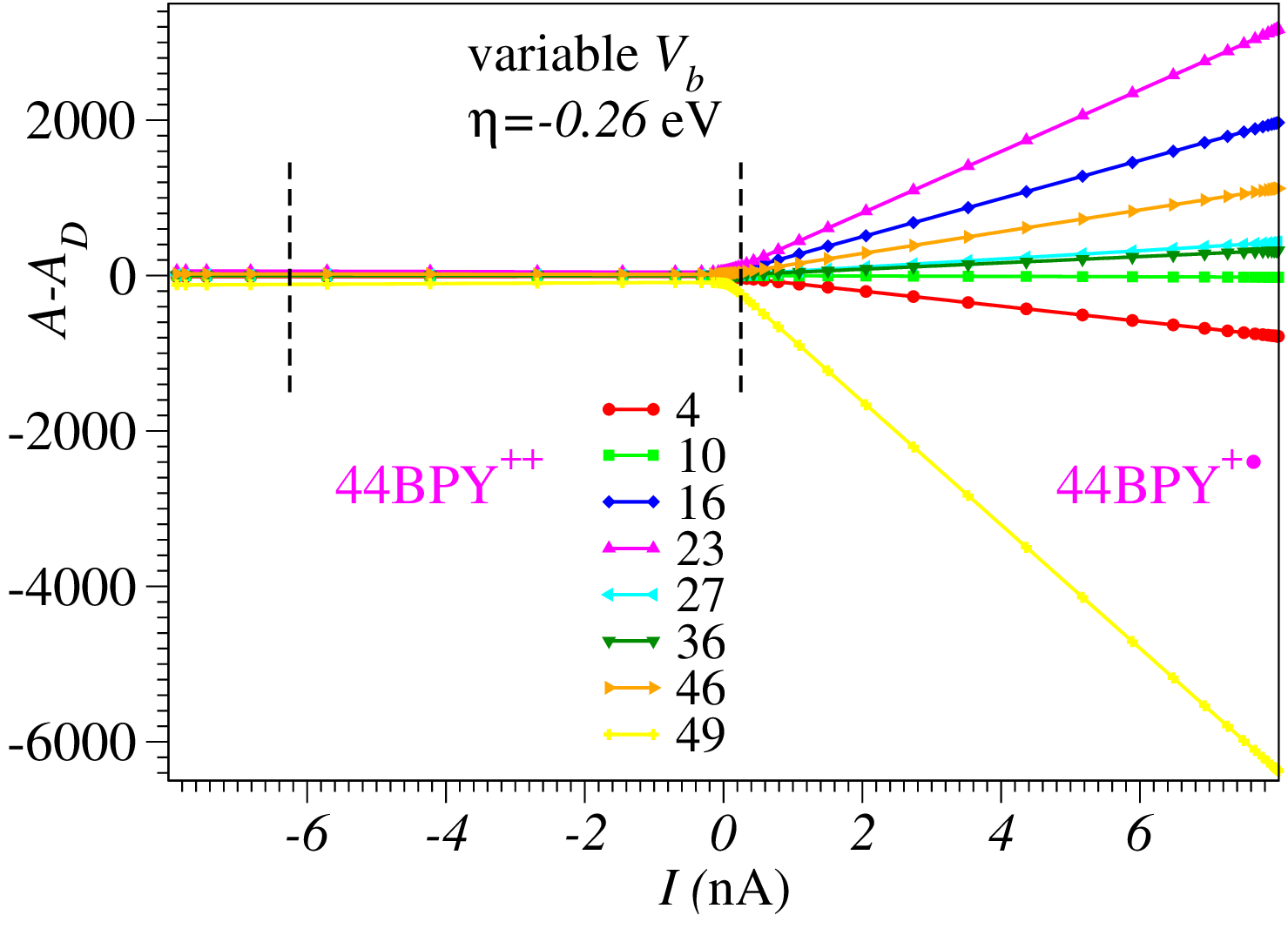}}
\caption{Dependence on the current $I$ of the changes in frequencies and Raman scattering activities (in {\AA}$^4$/a.m.u.) in variable bias mode
for several representative vibrational modes given in the legend.
The $V_b$-range sampled in experiment \cite{Wandlowski:08} is indicated by the two black vertical dashed lines.}
\label{fig:freq-A-vs-I-vary-V}
\end{figure}

\begin{figure}[h!]
$ $\\[5ex]
\centerline{\includegraphics[width=0.4\textwidth,angle=0]{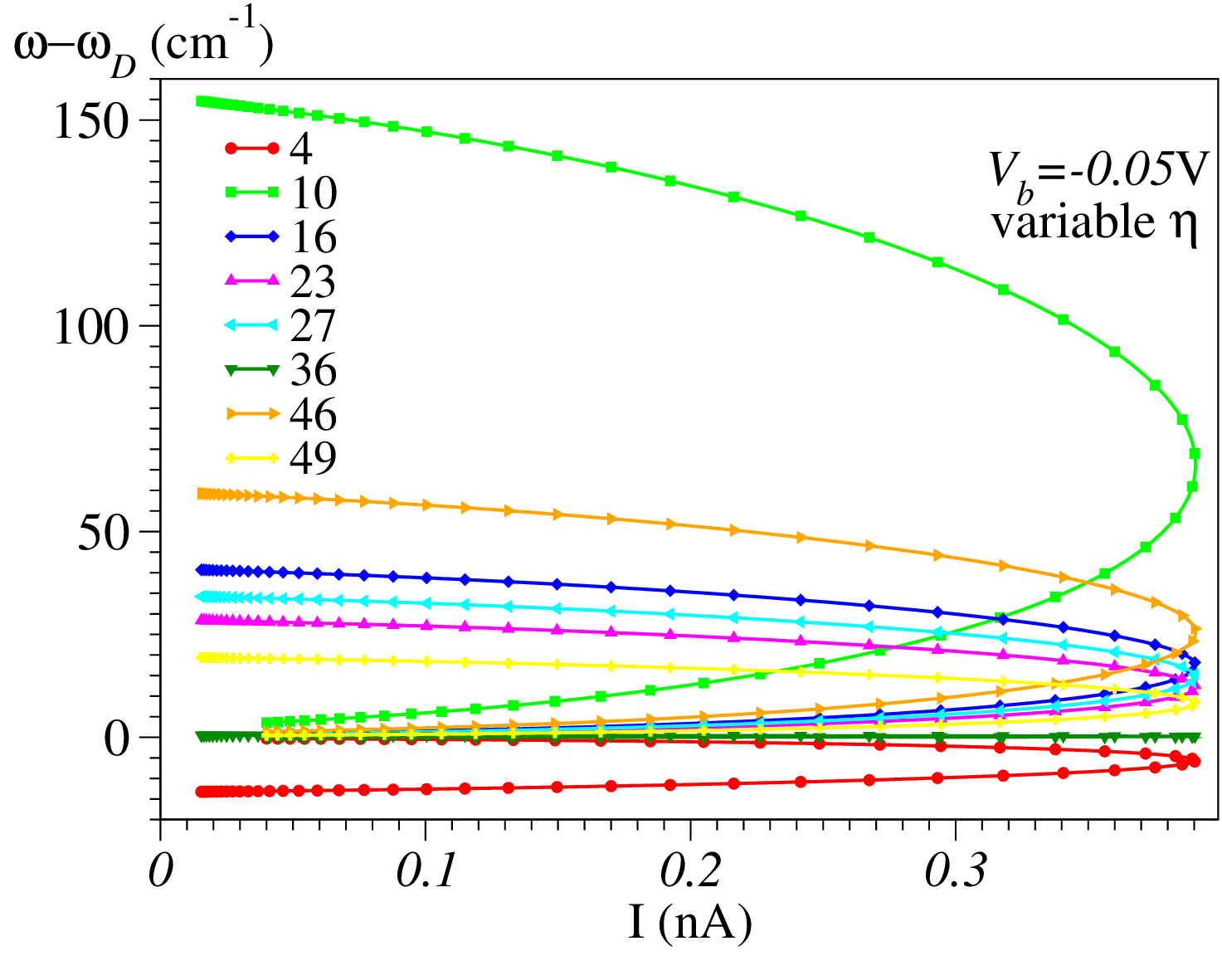}}
\centerline{\includegraphics[width=0.4\textwidth,angle=0]{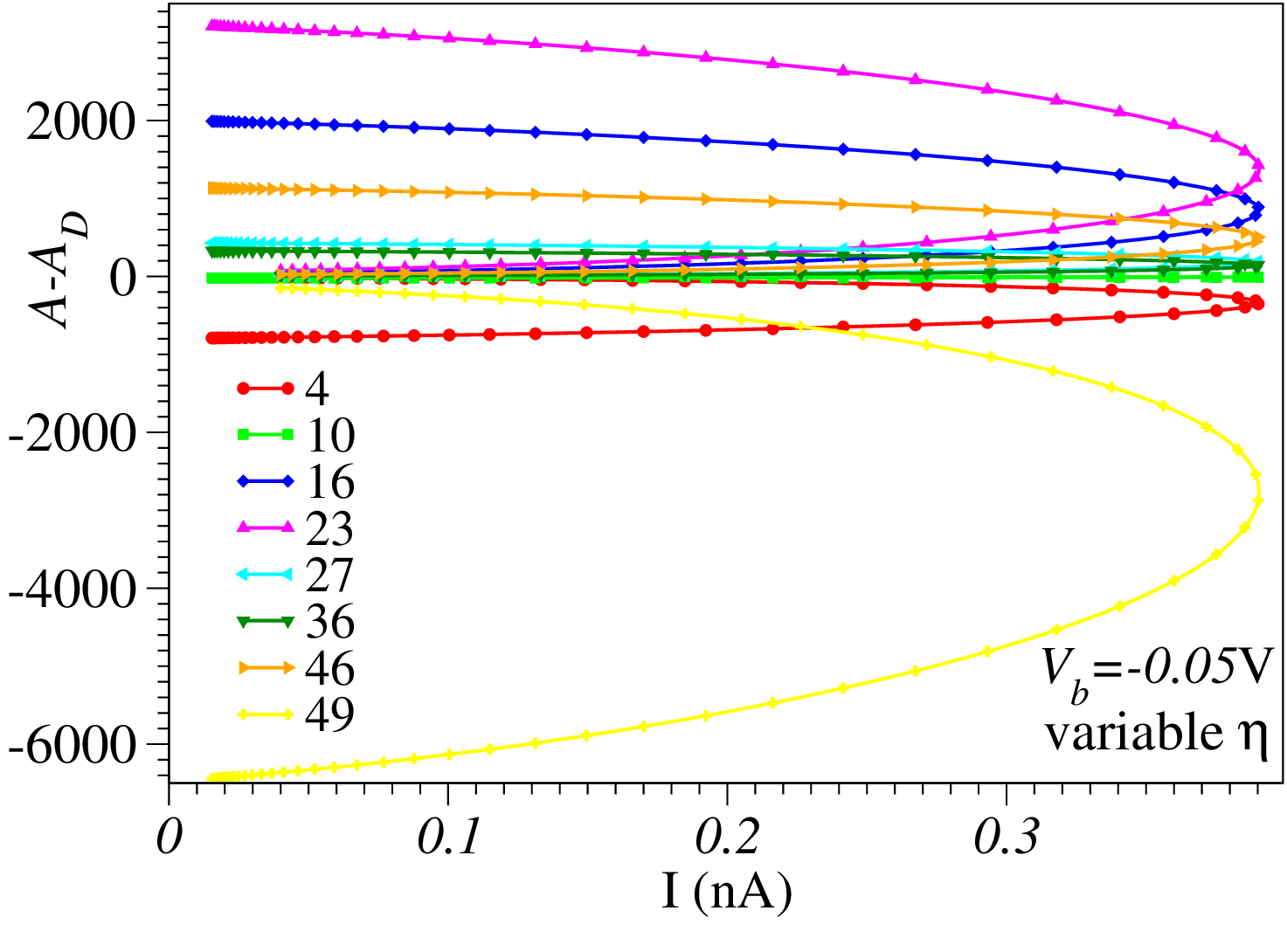}}
\caption{Dependence on the current $I$ of the changes in frequencies and Raman scattering activities
(in {\AA}$^4$/a.m.u.) in constant bias mode (variable $\eta$) for
several representative vibrational modes given in the legend.}
\label{fig:freq-A-vs-I-vary-eta}
\end{figure}

In an EC-STM setup similar to that of Ref.~\citenum{Wandlowski:08},
the electromagnetic (EM) enhancement occurs in a tiny region near the STM tip,
and the charge transport mainly proceeds through the single (physisorbed) molecule which is closest to it.
Still, other molecules nearby may also feel a significant electromagnetic enhancement and
may also experience LUMO energy shifts driven by more or less similar local
$V_b$ and $\eta$ values
[\emph{cf.}~Eq.~(S1)]; 
the spatial potential profile
is hard to control/determine. So, the recorded SERS dependence on $\eta$ (and $V_b$)
may not be (entirely) due to the (single) molecule mediating the charge transport.
Therefore, correlating the measured SERS with the currents measured by varying $\eta$ (or $V_b$)
rather than with $\eta$ (or $V_b$) is preferable; it would be the most clear indication that
the Raman signals come from the particular molecules inside the nanogap
that are responsible for the current between the electrodes.
Results in this form are depicted in
{\figurename}\ref{fig:freq-A-vs-I-vary-V} and
{\figurename}\ref{fig:freq-A-vs-I-vary-eta}.
They may be more useful to experimentalists than
{\figurename}\ref{fig:freq-A-vs-V} and {\figurename}\ref{fig:freq-A-vs-eta}.
\section{Remarks on experimental challenges}
\label{sec:challenges}
\ib{
Let us briefly motivate why, in the above presentation, we have 
(i) considered transport data for molecular junctions placed in solvent
and (ii) referred to SERS as \emph{the} technique 
to reveal the substantial dependence on the molecular charge 
of the vibrational properties predicted by our calculations.

(i) Given the impossibility of achieving a 
substantial reduction of the molecular species (significant change in the molecular charge)
by varying the source-drain bias $V_b$, an efficient molecular orbital gating 
appears to be irreplaceable. 
The molecular junctions studied in ref.~\citenum{Reed:09} exhibit
the most substantial orbital gating effect known to date in ``dry'' molecular electronics.
We checked by calculations similar to 
those presented above that, for those junctions, changes in molecular charges
do not exceed a few percent. This fact can actually be understood intuitively by examining, 
\emph{e.g.}, Fig.~S7 of the supplementary information of ref.~\citenum{Reed:09}, 
which shows curves that do not exhibit any peak; 
a substantial reduction (oxidation) can only be obtained if 
the gate potential $V_G$ utilized in experiments \cite{Reed:09} 
sample sufficiently extended portions of the $I-V_G$ transfer characteristics 
comprising a maximum (which is the counterpart of the maximum of the $I-\eta$-curve shown in {\figurename}\ref{fig:igna}). 

In principle, with a molecular orbital gating in ``dry'' environment substantially improved well beyond the present
achievements, \cite{Reed:09} changes in molecular vibrational properties comparable to those presented above 
could also occur in nanojunctions placed in vacuum, as demonstrated by the data in Table S3.
But, according to the state-of-the-art in the field,
a(n almost) complete reduction of the molecular species (changing the molecular charge by an entire electron)
can only be reached via the efficient electrolyte gating discussed in the present paper.

(ii) It is hard to conceive that IETS (inelastic electron tunneling spectroscopy) 
\cite{Ho:98,Ho:00,Ho:02} can be utilized
for the present purpose. Rather than SERS, IETS is the choice 
of experimentalists for revealing vibrational effects in the charge transport through molecular junctions,
\cite{Smit:02,Kushmerick:04,Reed:04a,Kiguchi:07,Hihath:08,Kiguchi:08,Reed:09,Scheer:11}
However, it requires cryogenic temperatures. Obviously, such conditions are totally inappropriate for 
molecular junctions immersed in solvents. 
The fact that SERS can be applied at room temperature 
\cite{Tao:06e,Tao:07a,Natelson:08a,Ioffe:08,Liu:11,Wandlowski:11c,Konishi:13a} is 
an important reason for advocating this method.

The experimental demonstration of the SERS-related effects under full electrochemical control discussed above 
is still pending, which may be related to certain experimental challenges, which we briefly mention below. 

An obvious challenge is the strength of the SERS signal. 
Valuable insight on the SERS enhancement factor can be gained from 
3D-FDTD (three-dimensional finite-difference time-domain) theoretical simulations, 
as shown in earlier work.\cite{Tao:06e,Wandlowski:11b,Wandlowski:11c,Konishi:13a}
In the absence of any experimental SERS information on the EC-STM junctions 
envisaged, a FTDT simulation would be too speculative and will not be attempted below.
In view of its well-known critical dependence on the incident light polarization, 
surface morphology, nanogap configuration and size,
an estimate of the SERS intensities can be realistic only if a sufficiently detailed experimental 
characterization is available.
Still, we think that the following pieces of experimental work on related nanosystems can be taken 
as positive signals that a combined transport-SERS study in EC-STM junctions like those considered in this paper 
is feasible: the extraordinary enhancement of the
Raman scattering from pyridine by shell-isolated gold nanoparticles, \cite{Wandlowski:11b}
the application of \emph{in situ} shell-isolated nanoparticle SERS
(SHINERS)\cite{Tian:10} under electrochemical condition for an interfacial redox reaction 
using the same (viologen) molecule as presently considered,\cite{Wandlowski:11c}
and the combined SERS and conductance measurements under the same experimental conditions 
for scanning tunneling microscope break junctions
of 44BPY placed in aqueous solution.\cite{Konishi:13a} In the last case,\cite{Konishi:13a}
a Raman enhancement factor of $3.8\times 10^8$ has been obtained. 
In view of the similarity between the presently considered system and that of ref.~\citenum{Konishi:13a},
we do not expect a dramatic reduction of the SERS intensity preventing observability.

Implementing SERS on single-molecule junctions \cite{Natelson:08a} represented an important advance in molecular electronics. The so-called ``fishing mode'' STM is the key for Raman signal accumulation. A shift of the Raman signal was observed and related to the stress applied to the molecule. 
Likewise, IETS (inelastic electron tunneling spectroscopy) studies 
on single-molecule junctions also indicated 
some shifts in vibrational modes under stress. 
So, stress-driven shifts could be a significant noise factor for the $\eta$- and $V_b$-dependent Raman signals. However, while this may be an issue \emph{in general}, 
and further joint experimental and theoretical studies are needed to clarify it, 
we believe that stress-driven shifts are less important for the \emph{specific} 
case considered in this paper. As noted in the {\si}, the viologen core (44BPY) 
is not directly contacted to (gold) electrodes in the molecular junctions 
used in experiment, but rather via alkyl linkers. These linkers 
mitigate the electrodes' impact on the 44BPY core. Therefore, we do not expect a
substantial impact on the intra-core vibrational modes similar to that caused by 
the change in the molecular charge considered above.
Related to this issue, 
it is noteworthy that even for the smaller 44BPY molecule (\emph{i.e.}, the molecule without alkyl(+thiol) linkers) 
vibration frequencies computed with gold atoms attached at the two ends of the molecule 
and only at one end are almost equal; see Table S1 of ref.~\citenum{Konishi:13a}.  

Obviously, neat curves like those presented in 
{\figurename}~\ref{fig:freq-A-vs-eta} and \ref{fig:freq-A-vs-I-vary-eta} 
represent a highly idealized description.
In real measurements, one should rather expect blurry curves, similar to those of previous reports.
\cite{Natelson:08a,Natelson:08c,Ioffe:08,Liu:11,Konishi:13a} 
In addition to the (albeit perhaps less pronounced) stress-driven shifts 
mentioned above, other deleterious effects such as heating \cite{Tao:06f,Tao:07b,Galperin:07,Ioffe:08,Natelson:08c,Thoss:09} and photocurrents 
\cite{Scheer:07,Natelson:08b,Pauly:08a,Ittah:09,Nitzan:11} are possible sources of noise. But just because such effects 
were also present in earlier successful SERS-transport studies, 
there is no special reason to assume that they are so strong to prevent observability
in the presently considered case. 
}
\section{Conclusion}
\label{sec-conclusion}
To conclude, in this paper we have theoretically demonstrated that
vibrational frequencies and Raman scattering
intensities of a redox molecule embedded in an EC-STM
single-molecule
junction, like those already fabricated,\cite{Wandlowski:08}
can be efficiently controlled
especially
by varying the overpotential (``gate'' potential).
As widely accepted,
for SERS observability,
the electromagnetic enhancement is decisive
although the chemical
(charge transfer)
enhancement \cite{Lombardi:08} may also play a role.\cite{Schatz:06}
The present results indicate that,
in addition to these,
the charge state may also be significant.
The information that can be gathered by SERS
goes beyond the chemical identity and valuable structural information
of the wired molecule and contacts;
the charge state of a molecule can also be probed via SERS.
Table S1 
shows that differences
in Raman intensities of the dicationic and cationic species
can be of several orders of magnitude.
\emph{E.g.}, reduction yields the Raman activity of mode 23 is enhanced by a
factor of $\sim$125 while for mode 4 it is diminished by a factor $\sim$640.
Using other solvents, the Raman activity enhancement can even be much larger 
(almost $10^4$), as illustrated by the results of Table S2 for modes 10 and 14 in benzene.
This is an effect much stronger than previously
reported in two-terminal setups without solvents.\cite{Mirjani:12}

Therefore, in spite of nontrivial experimental challenges mentioned above, 
we are confident that correlating transport data
and vibrational information acquired by SERS
from current carrying single-molecule junctions under electrochemical control
will open new avenues of research that are not accessible for nanotransport
in two-terminal setup. Practical molecular electronics and fundamental science
can be main beneficiaries.
%
\section*{Acknowledgments}
Financial support provided by the Deu\-tsche For\-schungs\-ge\-mein\-schaft
(grant BA 1799/2-1) is gratefully acknowledged.\\
%
\renewcommand\refname{Notes and references}
\footnotesize{
\bibliographystyle{rsc}
\providecommand*{\mcitethebibliography}{\thebibliography}
\csname @ifundefined\endcsname{endmcitethebibliography}
{\let\endmcitethebibliography\endthebibliography}{}

\end{document}